\documentclass[usenatbib,hyperpdf]{mn2e}
\usepackage{graphicx}

\def\ni{\noindent}                                    %No indent%
\def\etal{et\thinspace al.\thinspace}                    %et al.%
\def\starlight{\textsc{starlight}}                    %Starlight%
\newcommand{\Ha}{\ifmmode {\rm H}\alpha \else H$\alpha$\fi}
\newcommand{\Hb}{\ifmmode {\rm H}\beta \else H$\beta$\fi}
\newcommand{\oiii}{\ifmmode [\rm{O}\,\textsc{iii}] \else [O~{\sc iii}]\fi}
\newcommand{\nii}{\ifmmode [\rm{N}\,\textsc{ii}] \else [N~{\sc ii}]\fi}
\newcommand{\Oiii}{[O~{\sc iii}]$\lambda$5007}
\newcommand{\Nii}{[N~{\sc ii}]$\lambda$6584}
\newcommand{\Oii}{[O~{\sc ii}]$\lambda$3727}
\title[The evolution of the mass-metallicity relation in SDSS
galaxies]{The evolution of the mass-metallicity relation in SDSS
  galaxies uncovered by astropaleontology}

\author[Vale Asari \etal]
{N. Vale Asari$^{1,2}$,
  G. Stasi\'nska$^{2}$,
  R. Cid Fernandes$^{1}$,
  J. M. Gomes$^{1,3}$
  \newauthor
  M. Schlickmann$^{1}$,
  A. Mateus$^{4}$,
  W. Schoenell$^{1}$
  (the SEAGal collaboration)\thanks{Semi-Empirical Analysis of Galaxies} \\
  $^{1}$Departamento de F\'{\i}sica - CFM - Universidade Federal de
  Santa Catarina, Florian\'opolis, SC, Brazil\\
  $^{2}$LUTH, Observatoire de Paris, CNRS, Universit\'e Paris
  Diderot; Place Jules Janssen 92190 Meudon, France\\
  $^{3}$GEPI, Observatoire de Paris, CNRS, Universit\'e Paris
  Diderot; Place Jules Janssen 92190 Meudon, France\\
  $^{4}$Instituto de Astronomia, Geof\'{\i}sica e Ci\^encias
  Atmosf\'ericas, Universidade de S\~ao Paulo, S\~ao Paulo, SP,
  Brazil
}

\begin{document}

\maketitle

%***************************************************************%
%                                                               %
%                           Abstract                            %
%                                                               %
%***************************************************************%

\begin{abstract}
  We have obtained the mass-metallicity ($M$--$Z$) relation at
  different lookback times for the {\em same set of galaxies} from the
  Sloan Digital Sky Survey, using the stellar metallicities estimated
  with our spectral synthesis code \starlight. We have found that this
  relation steepens and spans a wider range in both mass and
  metallicity at higher redshifts. We have modeled the time evolution
  of stellar metallicity with a closed-box chemical evolution model,
  for galaxies of different types and masses. Our results suggest that
  the $M$--$Z$ relation for galaxies with present-day stellar masses
  down to $10^{10} M_\odot$ is mainly driven by the history of star
  formation history and not by inflows or outflows.
\end{abstract}


%***************************************************************%
%                                                               %
%                        Introduction                           %
%                                                               %
%***************************************************************%

\section{Introduction}
\label{sec:Introduction}

Since the study by \citet{Lequeux_etal_1979}, who found a
luminosity-metallicity relation for irregular galaxies, many papers
have reported the existence of a luminosity-metallicity or
mass-metallicity ($M$--$Z$) relation for all kinds of galaxies: The
more massive galaxies are also the ones with more metal-rich
interstellar medium (ISM).

Considerable work has also been devoted to explaining the $M$--$Z$
relation. For instance, \citet{Tremonti_etal_2004} find that outflows
play an important role in shaping the $M$--$Z$ relation, as the weaker
gravitation potential well of less massive galaxies makes them more
prone to lose enriched gas via galactic winds or supernovae
explosions.  \citet{Finlator_Dave_2008} claim that inflows of pristine
gas could also be a sound explanation, as the same amount of gas
falling into galaxies would have a greater impact on the chemical
composition of less massive ones.  \citet{Koppen_Weidner_Kroupa_2007}
advocate the role of an \emph{integrated} initial mass function: Massive
galaxies comprise more massive clusters which may contain more massive
stars that enrich the ISM faster.

The $M$--$Z$ relation derived from emission lines has been found to
change with redshift \citep{Savaglio_etal_2005,
  Maiolino_etal_2008}. The comparison between $M$--$Z$ at low and high
redshifts is, however, not straightforward. Different populations of
galaxies live at different redshifts, and this alone can affect the
derivation of the nebular metallicity, even if one were to apply the
same calibrations for both samples (see \citealp{Stasinska_2008}).
Nevertheless, at least qualitatively, the observed results are in
agreement with what is expected for the evolution of galaxies.

This brings us to another concern in $M$--$Z$ studies: the bias in the
derivation of the nebular metallicity from emission lines.  The
evolution of the mass-metallicity relation is usually studied in terms
of the abundance of oxygen in the ISM gas.  As known (see e.g.
\citealp{Kewley_Ellison_2008}), the measurement of nebular abundances
is very dependent on the method and calibrations used.

With our stellar population synthesis code \starlight, we can
determine the stellar metallicities, $Z_\star$, and the total masses
in stars, $M_\star$, at different epochs for any given galaxy
\citep{CidFernandes_etal_2007_SEAGal5}. We can thus follow the
evolution of the $M$--$Z$ relation \emph{for the same set of galaxies}
at different redshifts. \citet{Panter_etal_2008} used similar
techniques and discussed the nebular metallicities with respect to
stellar metallicities.  In this study, we rely on stellar
metallicities alone because, although they have their own problems,
they are free from the biases affecting nebular abundance
determinations.  The use of $Z_\star$ also allows us to explore more
massive galaxies, for which the nebular metallicity estimates are not
possible, such as AGN hosts and galaxies which have stopped forming
stars.

Since \starlight\ allows the determination of the star formation
histories (SFHs), we can go one step further and construct a simple
chemical evolution model using these SFHs and see under what
conditions it reproduces the observed time evolution of $Z_\star$.
\citet{Savaglio_etal_2005} made a similar study, but using a
theoretical SFH with e-folding time, related to galaxy mass, and for
nebular metallicity instead of $Z_\star$. Here we use the SFH obtained
{\em directly} for each galaxy.

%***************************************************************%
%                                                               %
%                            Data                               %
%                                                               %
%***************************************************************%

\section{Data}
\label{sec:Data}

%---------------------------- Figure ----------------------------
\begin{figure}
   \includegraphics[width=0.5\textwidth, bb=40 525 405 695]{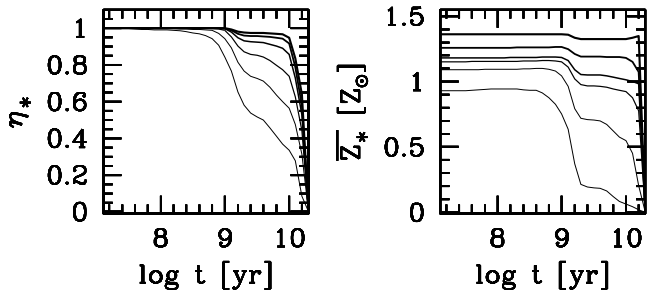}
   \caption{Histories of our parent sample in stellar mass
     bins. Thicker lines are used for more massive bins. Panels show
     the median curve for time evolution of stellar mass evolution
     (left; see Eq.~\ref{eq:MAH}) and metallicity evolution (right).}
\label{fig:Histories}
\end{figure}
%---------------------------- Figure ----------------------------

Our parent sample is composed of all objects spectroscopically
classified as galaxies from 5$^{\rm th}$ Data Release of the Sloan
Digital Sky Survey (SDSS)\citep{Adelman-McCarthy_etal_2007}. We
exclude duplicate observations, and impose the following selection
criteria: $14.5 \le m_r \le 17.77$ (from the Main Galaxy Sample
definition), $z$-band aperture covering factor $> 20\%$ (to reduce
aperture effects and avoid intragalactic sources), $S/N \ge 10$ at
4750 \AA\ (to provide reasonable stellar population fits), and a
narrow range in redshift $|z - 0.1| < 0.015$ (to be able to transform
stellar ages into redshift in a simple way).  The central redshift of
0.1 was chosen to maximize the number of objects and thus ensure
reliable statistics over the whole range of ages considered in the
analysis.  We adopt a $H_0 = 70 {\rm \, km \, s^{-1} \, Mpc^{-1}}$,
$\Omega_M = 0.3$ and $\Omega_\Lambda = 0.7$ cosmology where
appropriate.  There are 82662 objects in the resulting sample.

We use the same data processing as outlined in
\citet{CidFernandes_etal_2005_SEAGal1} and
\citet{Mateus_etal_2006_SEAGal2}. Our code \starlight\ recovers the
stellar population content of a galaxy by fitting a pixel-by-pixel
model to the spectral continuum. This model is a linear combination of
150 simple stellar populations (SSP) extracted from \citet[hereafter
BC03]{Bruzual_Charlot_2003_BC03} with ages 1 Myr $\le t_\star \le$ 18
Gyr, and metallicities $0.005 \le Z/Z_\odot \le 2.5$, as in
\citet{CidFernandes_etal_2007_SEAGal5}.  Emission lines are then
measured in the residual spectra, which reduces the contamination by
stellar absorption features.

We divide our parent sample into three sub-samples: star-forming (SF),
retired (R) and Seyfert (S) galaxies. For both the SF and S samples,
we select objects with $S/N > 3$ in the \Oiii, \Hb, \Nii\ and \Ha\
emission lines, i.e., all the lines involved in the
\citet*[BPT]{Baldwin_Phillips_Terlevich_1981_BPT} diagram. The SF
sample is composed by the objects below the line proposed by
\citet{Stasinska_etal_2006_SEAGal3} to separate SF galaxies from AGN
hosts. The Seyfert sample is defined as the galaxies with $\log
(\nii/\Ha) > -0.3$ and above the line defined by
\citet{Schlickmann_2008} in the BPT diagram to separate Seyferts from
LINER-like galaxies: $\log (\oiii/\Hb) = 0.90 \log (\nii/\Ha) + 0.48$.

The R sample is composed of galaxies with little evidence of
either nuclear or star-forming activity, which includes passive
galaxies, galaxies with very weak emission lines and even some which
would be traditionally classified as LINERs (see
\citealp{Stasinska_etal_2008}). Therefore, we impose the following
cuts for the R sample: $S/N < 3$ in the 4 BPT lines and
\Oii\ (``passive''); or $S/N < 3$ in \oiii\ and/or \Hb, but $S/N \ge
3$ in \nii\ and \Ha, and $\log (\nii/\Ha) > -0.15$ (``weak emission
lines''); or $S/N > 3$ in the 4 BPT lines, $\log (\nii/\Ha) > -0.15$,
below the line defined by \citet{Schlickmann_2008}, and \Ha\ emission
predicted by old stellar populations equal or greater than the
observed value (``retired mimicking LINERs'')

We have also divided our sample and sub-samples into six present-day
stellar mass bins centered in $\log M_\star/M_\odot = 10.0$ (A), 10.3
(B), 10.6 (C), 10.9 (D), 11.2 (E) and 11.5 (F), each one 0.30 dex
wide. In the same vein as our previous works
\citep{CidFernandes_etal_2007_SEAGal5, Asari_etal_2007}, defining bins
allows us to alleviate the problem of the mass completeness of the
sample. We consider narrow mass bins in order to have very similar
galaxies inside each bin. Although the starting sample contained a
small proportion of galaxies with masses smaller than $10^{10}
M_\odot$, we do not consider them in the current analysis. Indeed, due
to the redshift cut we imposed, only the most luminous low-mass
objects are present and they are not representative of the bulk of
galaxies of their mass range.

%***************************************************************%
%                                                               %
%                       Defining stuff                          %
%                                                               %
%***************************************************************%

\section{Galaxy histories in mass bins}
\label{sec:definitions}

%---------------------------- Figure ----------------------------
\begin{figure*}
   \centering
   \includegraphics[width=0.6\textwidth, bb=170 480 592 700]{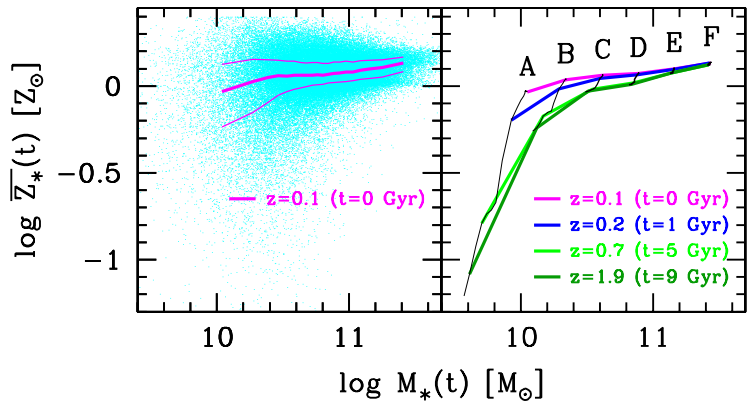}
   \caption{Left: Present-day mass-metallicity relation for all the
     galaxies in our sample. The thick line marks the median and thin
     lines the quartiles. Right: Mass-metallicity evolution for each
     of our present-day stellar mass bins A--F (black). From top to
     bottom, bold lines mark the mass-metallicity relation for
     redshifts $z = 0.1$ ($t = 0$ Gyr), 0.2 (1 Gyr), 0.7 (5 Gyr) and
     1.9 (9 Gyr).}
\label{fig:MxZ_evolution}
\end{figure*}
%---------------------------- Figure ----------------------------

As explained in \citet{CidFernandes_etal_2007_SEAGal5}, \starlight\
recovers the fraction $x_j$ that each SSP $j$ contributes to the total
light of a galaxy in the spectral range covered by the SDSS. We can
translate $x_j$ into the mass fraction {\em presently locked} into
stars, $\mu_j$, or the fraction of mass {\em ever converted} into
stars, $\mu_j^c$.

The stellar mass history can be followed as a function of
lookback time $t$ by summing up the converted-mass fractions:

\begin{equation}
\label{eq:MAH}
\eta_\star(t) =
     \sum_{t_j > t} \mu^c_j {\rm .}
\end{equation}

The mean stellar metallicity at a lookback time $t$ is defined as the
total mass in metals locked in stars divided by the total stellar mass
at a given time:

\begin{equation}
\label{eq:zeta}
\overline{Z_\star}(t) = 
\sum_{t_j > t} \mu_j Z_{\star,j} {\rm .}
\end{equation}

Another way to look at the metallicity evolution would be to consider
the average metallicity of all the stars born at the same epoch, for
each value of $t$. This, however, is a much less robust quantity, and
we found that SSP bases at hand nowadays do not allow such detailed
definition.

Figure~\ref{fig:Histories} shows the median values of $\eta_\star$ and
$\overline{Z_\star}$ versus $t$ for mass bins A--F of our sample. As
previous studies have already shown
(e.g. \citealp{CidFernandes_etal_2007_SEAGal5, Panter_etal_2008}, for
SF galaxies only -- here we extend the study for all galaxies), there
is a clear {\em downsizing} effect: The less massive galaxies are
slower in converting their mass into stars and in building up their
metal content.

%***************************************************************%
%                                                               %
%                             MxZ                               %
%                                                               %
%***************************************************************%

\section{The observed $M$--$Z$ evolution}
%\section{Mass-metallicity empirical evolution}
\label{sec:MxZ}

Figure~\ref{fig:MxZ_evolution} shows the mass-metallicity relation.
The left panel shows the {\em present-day} $M$--$Z$ relation for all
the galaxies in our sample.  The right panel depicts the {\em
  evolution} of the $M$--$Z$ relation for the galaxies in mass bins
A--F (black lines): $\overline{Z_\star}(t)$ vs. $M_\star(t)$, where
$M_\star(t)$ is the stellar mass integrated from the epoch of galaxy
formation until a lookback time $t$, assuming that the stars always
pertain to the galaxies where they were born.  The resulting $M$--$Z$
relations for different redshifts are plotted in bold lines.  One can
see that the mass-metallicity relation evolves.

As lookback time increases, the $M$--$Z$ relation steepens and covers
a larger range of values. As investigated by \citet{Panter_etal_2008},
using stellar modelling codes other than that of BC03 changes the
values of $M_\star(t)$ and $\overline{Z_\star}(t)$, but the general
behaviour seen in the present paper as well as in
\citet{Panter_etal_2008} remains.  One should note that the $M$--$Z$
relation and evolution is significantly sample-dependent, which makes
difficult any quantitative comparison between works by different
authors.

%***************************************************************%
%                                                               %
%                      Chemical Evolution                       %
%                                                               %
%***************************************************************%

\section{A simple chemical evolution model}
\label{sec:ChemEvol}

We now investigate the behaviour of the
$M_\star$--$\overline{Z_\star}$ relations shown in
Fig. \ref{fig:MxZ_evolution} with the help of a simple model of
chemical evolution of galaxies.  We assume a closed box with initial
metallicity equal to zero, and use an instantaneous recycling
approximation and a constant yield. The mass-weighted average mass
fraction of metals at lookback time $t$ is then given by
(\citealp{Edmunds_1990}, eq. 35):

\begin{equation}
\label{eq:Ze}
\overline{Z_\star}(t) = y - y \frac{ f(t) \ln(1/f(t)) }{ (1-f(t)) },
\end{equation}

\ni where $f(t)$ is the gas mass fraction and $y$ is the ratio of the
mass of metals released per stellar generation to the total stellar
mass locked in remnants.  The gas mass fraction is given by:

\begin{equation}
\label{eq:ft1}
f(t) = 1 - \left( \frac{1 - R}{M_T} \right)
           \int_{t}^{t_{\rm o}} {\rm SFR}(t^\prime) \, {\rm
             d}t^\prime,
\end{equation}

\ni where $M_T$ is the total mass of the galaxy (stars and gas), SFR
is the star formation rate, $R$ is the returned mass fraction due to
stellar winds and supernovae and $t_{\rm o}$ is the age of the oldest
SSP in the base. The star formation rate SFR$(t)$ can be obtained from
the results of the \starlight\ synthesis:

\begin{equation}
\label{eq:SFR}
{\rm SFR}(t) dt = 
       M^c_\star \mu^c_\star(t),
\end{equation}

\ni where ${M^c_\star}$ is the total mass converted into stars
(${M^c_\star} = {M_\star}/(1 - R)$).  Using Eqs.  \ref{eq:MAH} and
\ref{eq:SFR}, $f(t)$ becomes simply:

\begin{equation}
\label{eq:ft2}
f(t) = 1 - (1 - f_{\rm now}) \eta_\star(t) {\rm ,}
\end{equation}

\ni where $f_{\rm now}$ is the present-day gas mass fraction.  Knowing
the value of $y$, one can use Eqs. \ref{eq:Ze} and \ref{eq:ft2} with
$\eta_\star(t)$ given by \starlight\ to find what value of $f_{\rm
  now}$ is needed to reproduce the observed evolution of the mean
stellar metallicity $\overline{Z_\star}(t)$.

Unfortunately, $y$ is not well known. The measured stellar
metallicities are mainly sensitive to the abundance of iron, and the
Fe/O ratio is expected to vary with time.  For our simple approach, we
assume that all the metals vary in lockstep and simply speak of the
``metal yield''. We determine an empirical value of $y$ by applying
our model to the highest mass bin of the SF sample, using the
Schmidt-Kennicutt law \citep{Kennicutt_1998}\footnote{We use the
  Schmidt-Kennicutt law for normal disk galaxies (bivariate
  least-squares regression in fig.~2 of \citealp{Kennicutt_1998}) for
  a \citet{Chabrier_2003} initial mass function, and including the
  contribution of helium in the gas mass fraction.}  to derive the
value of $f_{\rm now}$ from the extinction-corrected
\Ha\ luminosity. This procedure should give a reasonable value of the
metal yield, since it is for the highest mass bin that the closed box
model is expected to be the most relevant. We obtain $y=0.03$. This
value is in satisfactory agreement with the value ($0.04$) obtained
for a \citet{Chabrier_2003} stellar initial mass function by
multiplying the oxygen yield from \citet{Woosley_Weaver_1995} by the
Solar metal-to-oxygen mass fraction.

%---------------------------- Figure ----------------------------
\begin{figure*}
   \centering
   \includegraphics[width=0.95\textwidth, bb=30 445 550 710]{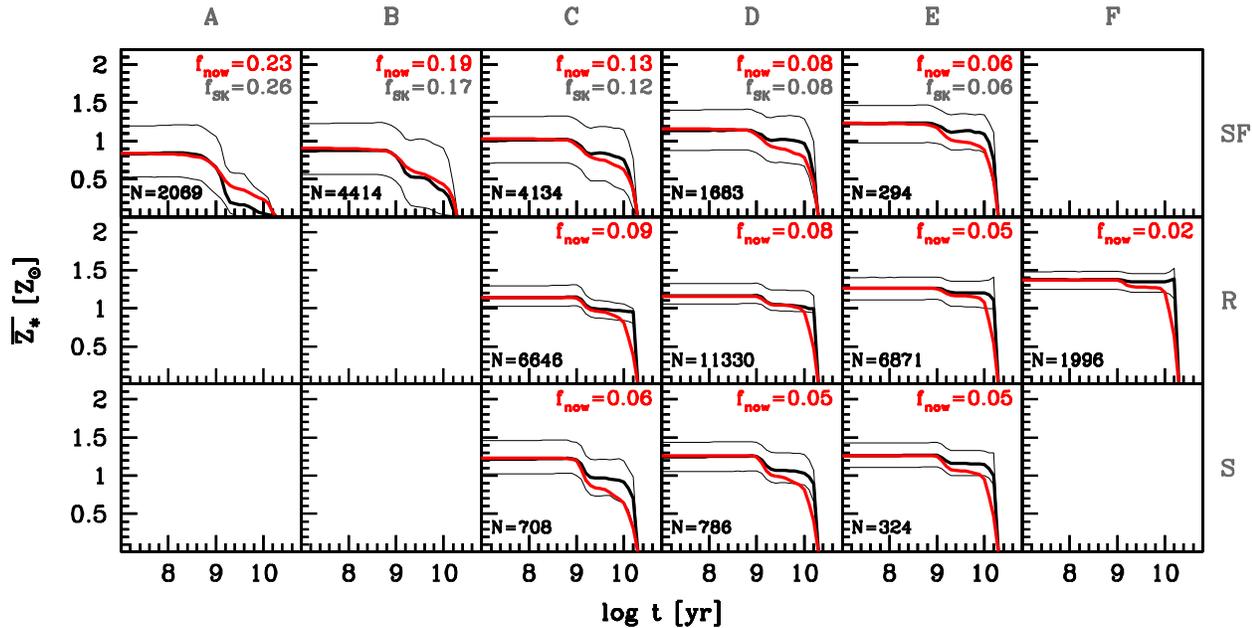}
   \caption{Chemical evolution in our samples of SF, R and S galaxies
     in the 0.30 dex-wide stellar mass bins, centered in $\log
     M_\star/M_\odot = 10.0$ (A), 10.3 (B), 10.6 (C), 10.9 (D), 11.2
     (E) and 11.5 (F). Each panel shows the median and quartiles of
     the evolution of $Z_\star$ as found by \starlight\ (black lines)
     and as obtained with the simple closed-box model (red lines).
     The value of $f_{\rm now}$ needed to reproduce the median
     present-day $Z_\star$ is indicated at the top right of each
     panel. For the SF sample, we also indicate the median of $f_{\rm
       now}$ as measured by the Schmidt-Kennicutt law ($f_{\rm SK}$).
     The yield $y$ was adjusted to reproduce bin E of the SF sample.
     The number of galaxies in each panel is indicated at the
     bottom. Panels were left empty if they contained less than 250
     galaxies.}
\label{fig:ChemEvol}
\end{figure*}
%---------------------------- Figure ----------------------------

We therefore adopt $y=0.03$ and apply our model to try to reproduce
the observed metallicity evolution for all the mass bins.
Figure~\ref{fig:ChemEvol} shows $\overline{Z_\star}(t)$ as a function
of $t$ for bins A--F of our SF, R and S galaxy samples.  The black
lines represent the median and quartiles of the observed
distributions. The red lines are the model results for the values of
$f_{\rm now}$ indicated at the top of each panel.  The models were
adjusted by eye to reproduce well at least the last 1 Gyr of the
observed $\overline{Z_\star}(t)$.  For the SF sample, we also indicate
the median value of $f_{\rm now}$ obtained from the Schmidt-Kennicutt
law.  This figure shows that this simple model can reproduce the
observed metallicity evolution reasonably well, given the crudeness of
our approach.  The discrepancies for ages above $10^9$ yr are at least
partly due to the known problems faced by spectral synthesis in this
range (\citealp{Gomes_2005}; \citealp{Koleva_etal_2008}). In
particular, the jumps in both $\eta_\star$ and $\overline{Z_\star}$ at
$\sim 1$ Gyr are artifacts related to the spectral libraries used in
BC03 \citep{CidFernandes_etal_2008}.  New evolutionary synthesis
models with more accurate stellar spectra should improve the recovery
of SFHs in this range.

It is worth of noting that $f_{\rm now}$ is at least qualitatively in
agreement with expectations: it is larger for SF galaxies than for
retired galaxies, and decreases with increasing $M_\star$.  As a
matter of fact, for all the mass bins of the SF sample $f_{\rm now}$
is compatible with the values derived from the Schmidt-Kennicutt law
(bin E is compatible by construction).  Seyferts also shows little
residual gas mass. The chemical evolution of Seyferts is (within our
resolution) similar to the evolution of non-Seyfert galaxies.

To sum up, for present-day stellar masses down to $10^{10} M_\odot$
and within the current limitations of astropaleontology, the closed
box model explains the present-day $M$--$Z$ relation and its evolution
quite well if one uses the SFR$(t)$ obtained \emph{directly} from the
spectral synthesis of each galaxy.  This is not to say that infall or
outflow do not play a role.  What we want to stress is that it is the
star formation history which is the main driver of the
mass-metallicity relation and its evolution.

%***************************************************************%
%                                                               %
%                          Summary                              %
%                                                               %
%***************************************************************%

\section{Summary and discussion}
\label{sec:Discussion}

Until recently, the $M$--$Z$ relation in galaxies had been mostly
studied using abundances derived from emission lines.  Exploiting the
\emph{stellar} metallicities obtained with our spectral synthesis code
\starlight, we have obtained the $M$--$Z$ relation at different
lookback times for the {\em same set of galaxies}. The use of stellar
metallicities, even if those are not very accurate, has several
merits. It allows one to probe metallicities at different epochs of a
galaxy evolution. It makes it possible to study galaxies in a larger
mass range. Galaxies without emission lines or galaxies with an active
nucleus do not have to be discarded. Finally, with stellar
metallicities, one avoids the biases that affect the statistical
methods used to derive metallicities from emission lines.

Our main results are the following.  We have found that the $M$--$Z$
relation steepens and spans a wider range in both mass and metallicity
at higher redshifts. The more massive galaxies show very little
evolution since a lookback time of 9 Gyr. This means that they evolved
very fast in a distant past, supporting the {\em downsizing} scenario.
This is in agreement with other studies of the build-up of metals in
SDSS galaxies as revealed by fossil records, e.g.,
\citet{CidFernandes_etal_2007_SEAGal5} and \citet{Panter_etal_2008}.

We have modeled the observed time evolution of the mean stellar
metallicity using a closed-box, instantaneous recycling chemical
evolution model, for galaxies of different types and masses. We find
that this model is compatible with the observations.  This suggests
that, in the mass range studied here (from $\log M_\star/M_\odot =
9.85$ to 11.65), the $M$--$Z$ relation is mainly driven by the history
of star formation and not by inflows or outflows.  By comparing the
nebular and stellar metallicities of SDSS galaxies,
\citet{Gallazzi_etal_2005} had argued that galaxies are not well
approximated by closed-box models and that winds may be important.
Our study suggests that those processes are not dominant, although
they may play a role in the {\em scatter} of the $M$--$Z$ relation in
our mass range.

Low-mass galaxies are not present in our sample because of our
selection in redshift. They would require a dedicated study to be
compared to the result found by \citet{Tremonti_etal_2004} and
\citet{Garnett_2002} that such galaxies do suffer strong outflows.  As
a matter of fact, Garnett finds that outflows are important for
galaxies with $B$-band magnitude $> -18$. Only 6 out of the 82662
objects in our sample have $g$-band magnitude $> -18$.  Using the
Schmidt-Kennicutt law to compute the effective yields for each galaxy
(i.e., the yield calculated from the observed nebular metallicities
assuming a closed box model), \citet{Tremonti_etal_2004} concluded
that galactic winds are ubiquitous and very effective in removing
metals from galaxies. This however concerns only galaxies with masses
up to $10^{10} M_\odot$, which are not present in our sample. Our work
emphasizes that there is a large range of galaxy masses -- almost two
decades -- where the closed box model seems to give a valid
representation of the chemical evolution of galaxies.

Our models are parametrized by the present-day gas mass fraction,
$f_{\rm now}$. We have found that the values of $f_{\rm now}$ that
allow us to reproduce the observed $\overline{Z_\star}(t)$ histories
are larger for less massive galaxies, in agreement with
expectations. For galaxies of similar masses, they are larger for
star-forming galaxies than for retired galaxies.  The values of
$f_{\rm now}$ we find for SF galaxies are compatible with those
obtained from the Schmidt-Kennicutt law.  This sound result supports
the validity of our interpretation. 

Within our uncertainties, Seyfert galaxies have as little gas as
retired galaxies.

%***************************************************************%
%                                                               %
%                       Acknowledgments                         %
%                                                               %
%***************************************************************%

\section*{ACKNOWLEDGMENTS}

The \starlight\ project is supported by the Brazilian agencies CNPq,
CAPES, FAPESP, by the France-Brazil CAPES-COFECUB program, and by
Observatoire de Paris. We thank Laerte Sodr\'e, Cristina Chiappini and
Nikos Prantzos for their helpful comments on the manuscript. We are
also grateful to the referee for interesting suggestions.

The Sloan Digital Sky Survey is a joint project of The University of
Chicago, Fermilab, the Institute for Advanced Study, the Japan
Participation Group, the Johns Hopkins University, the Los Alamos
National Laboratory, the Max-Planck-Institute for Astronomy, the
Max-Planck-Institute for Astrophysics, New Mexico State University,
Princeton University, the United States Naval Observatory, and the
University of Washington.  Funding for the project has been provided
by the Alfred P. Sloan Foundation, the Participating Institutions, the
National Aeronautics and Space Administration, the National Science
Foundation, the U.S. Department of Energy, the Japanese
Monbukagakusho, and the Max Planck Society.

%***************************************************************%
%                                                               %
%                          References                           %
%                                                               %
%***************************************************************%

%\begin{thebibliography}{}
%\end{thebibliography}

\bibliographystyle{mn2e}
\bibliography{bibliografia}

\end{document}